\documentclass[conference]{IEEEtran}
\usepackage{cite}
\usepackage{graphicx}

\usepackage{hyperref}
\hypersetup{colorlinks,allcolors=black}

\usepackage{amsmath}
\usepackage{caption}
\usepackage{subcaption}
\graphicspath{ {images/} }

\begin{document}


\title{Hybrid Intelligent Testing in Simulation-Based Verification}


\author
{\IEEEauthorblockN{Nyasha Masamba}
\IEEEauthorblockA{Faculty of Engineering\\Trustworthy Systems Laboratory\\
University of Bristol\\
United Kingdom\\
Email: nyasha.masamba@bristol.ac.uk }
\and
\IEEEauthorblockN{Kerstin Eder}
\IEEEauthorblockA{Faculty of Engineering\\Trustworthy Systems Laboratory\\
University of Bristol\\
United Kingdom\\
Email: kerstin.eder@bristol.ac.uk }
\and
\IEEEauthorblockN{Tim Blackmore}
\IEEEauthorblockA{Infineon Technologies\\Stoke Gifford\\
Bristol\\
United Kingdom\\
Email: tim.blackmore@infineon.com }
}
\maketitle


\begin{abstract}
\label{abstract}
Efficient and effective testing for simulation-based hardware verification is challenging. Using constrained random test generation, several millions of tests may be required to achieve coverage goals. The vast majority of tests do not contribute to coverage progress, yet they consume verification resources. In this paper, we propose a hybrid intelligent testing approach combining two methods that have previously been treated separately, namely Coverage-Directed Test Selection and Novelty-Driven Verification. Coverage-Directed Test Selection learns from coverage feedback to bias testing toward the most effective tests. Novelty-Driven Verification learns to identify and simulate stimuli that differ from previous stimuli, thereby reducing the number of simulations and increasing testing efficiency. We discuss the strengths and limitations of each method, and we show how our approach addresses each method's limitations, leading to hardware testing that is both efficient and effective.

\end{abstract}

\begin {IEEEkeywords}
Design Verification, Intelligent Testing, Coverage-Directed Test Selection, Novelty-Driven Verification, Machine Learning for Verification, CDS, CDG, EDA
\end{IEEEkeywords}

\section{Introduction}
\label{introduction}

Functional verification is the process of reconciling a design's functional specification with its implementation to gain confidence in the functional correctness of the design \cite{Bergeron2003}. Simulation-based verification is a type of functional verification which involves generating test stimuli, stressing the design under test (DUT) with those tests in simulation, and checking that the resulting DUT behaviour adheres to the specification. Deviations between DUT behaviour are flagged and investigated, as they could be the result of faults -- commonly referred to as bugs -- in the design logic. 

When stimuli exercise some given DUT functionality, the event is recorded in a coverage database. Coverage is an important measure of verification progress and test quality in simulation-based verification. Even modest sized DUTs have vast functionality that renders exhaustive verification impractical. Coverage helps the verification engineer to narrow down verification efforts to areas of the DUT deemed crucial for operation. By monitoring coverage, verification engineers know when sufficient testing has been done. Based on analysis of the coverage reports, verification engineers also know where to focus their efforts by biasing test generation to target uncovered areas. 

There are two hard challenges in simulation-based verification, namely generating effective test stimuli, and increasing verification efficiency. Effective tests are those that are legal with respect to the DUT specification, and useful at exercising important and hard-to-reach DUT functionality. Being efficient means reducing consumption of verification resources such as computation, EDA licences, simulation time and manual engineer effort. 

Constrained random test generation is currently the state-of-the-art test generation method in simulation-based verification. However, constrained random test generation can be inefficient because a large number of constrained random tests end up exploring the same functionality. Such inefficiencies lead to needless consumption of verification resources. Constrained random test generation tends to be an ineffective method for targeting coverage holes. Coverage holes are functional areas of the DUT that have been specifically identified for verification, but currently remain unverified. Manual, time-consuming constraint biasing by senior verification engineers is usually employed to overcome the ineffectiveness of constrained random testing.

Intelligent testing offers a solution for improving constrained random test generation. Intelligent testing refers to methods employing artificial intelligence (AI) \cite{RussellNorvig2010} to increase the efficacy and efficiency of constrained random test generation. Earlier intelligent testing approaches were based on AI techniques such as Bayesian Networks \cite{IBM2003} and Inductive Logic Programming \cite{Ederetal07}. These earlier approaches can be broadly categorised as Coverage-Directed Test Generation (CDG) methods \cite{Ioannides2012}. CDG methods generally use AI to discover relationships between coverage feedback and test generation constraints. The insights derived from CDG are used to increase effectiveness of generated stimuli by biasing test generation towards plugging coverage holes. However, CDG has historically required significant manual encoding of both DUT and domain knowledge into the AI model. Consequently, these earlier works on intelligent testing could only manage to experiment with relatively small DUTs, aside from replacing one form of manual work (writing constraints) with another (encoding knowledge into an AI model). 

An adaptation of CDG, coverage-directed test selection \cite{Masamba2022} introduced an approach for modelling relationships between functional coverage and test stimuli through supervised learning. Coverage-directed test selection (CDS) is ideal when a large set of constrained random test stimuli can be generated at little effort and cost. If significantly more verification resources are required to simulate those stimuli, then it is prudent to select effective stimuli from the large set, and then prioritise them for simulation. CDS focuses on biasing test selection towards the most effective tests. It uses supervised learning to extract constraints from labelled coverage feedback. Extracted constraints are subsequently used to select stimuli that have the highest probability of exercising targeted coverage holes. By automatically biasing test selection and simplifying the modelling process to a point where any binary classifier can be deployed, CDS reduces manual constraint writing while improving on CDG's manual encoding of knowledge into an AI model. 

Novelty-driven verification \cite{Blackmoreetal2021} utilises unsupervised novelty detection to identify novel stimuli from a large set of constrained random tests, then prioritises them for simulation. Similar to CDS, novelty-driven verification (NDV) is ideal for verification environments in which test generation is significantly cheaper than simulation. Unlike CDS, the NDV approach does not build a model relating functional coverage to test stimuli. Instead, an unsupervised model of similarity is constructed from previously simulated tests. The model is used to rank new simulation candidates, with preference given to the most dissimilar -- i.e., novel -- tests. The focus of NDV is to determine the order in which to simulate test stimuli such that functional coverage progress is accelerated due to redundancy reduction. Efficiency is realised in the form of reduced simulation, as high coverage is achieved using fewer test stimuli. However, the efficacy of NDV tests depends on the efficacy of the test biasing mechanism. This issue arises because of NDV's usage of novelty as a proxy for effectiveness: it is possible for stimuli to be novel, but not effective at targeting coverage holes. In light of this issue, CDS has the potential to enhance the efficacy of the NDV approach.

CDG and NDV have evolved separately in the literature to become the two most prominent categories of intelligent testing methods within simulation-based verification. We show that treating CDS and NDV as complementary methods, as suggested in \cite{LCWang2012}, has the potential to increase both the efficacy of stimuli, and overall verification efficiency. This results in higher verification productivity, accelerated coverage closure, and more time for verification engineers to hunt bugs in unexplored parts of the DUT. We combine CDS with NDV to create two types of hybrid intelligent testing frameworks. The first hybrid framework performs CDS and NDV sequentially at different stages of the verification process. This sequential mode of operation simulates the union of CDS and NDV test stimuli. The second hybrid framework performs CDS and NDV in parallel during the final stages of the verification process. This parallel mode of operation simulates the intersection of CDS and NDV stimuli. We refer to the novel intelligent testing approaches as the Unified Hybrid Approach, and the Intersected Hybrid Approach, respectively. To the best of our knowledge, this has not been done before. 

This paper is structured as follows. Section \ref{intelligent_testing} explains the objectives of intelligent testing, and then introduces formalised CDS and NDV theory. CDS and NDV are introduced as part of a hybrid intelligent testing framework in Section \ref{hybrid_approach}, where the Unified Hybrid Approach and the Intersected Hybrid Approach are presented in detail. In Section \ref{experiment}, the DUT is presented, along with experiment information. The experimental results are presented in Section \ref{results}. Section \ref{conclusion} concludes the paper by discussing implications and limitations of the research, along with planned future research.


\section{Intelligent Testing}
\label{intelligent_testing}

\begin{figure}[t]
    \centering
    \includegraphics[width=0.49\textwidth]{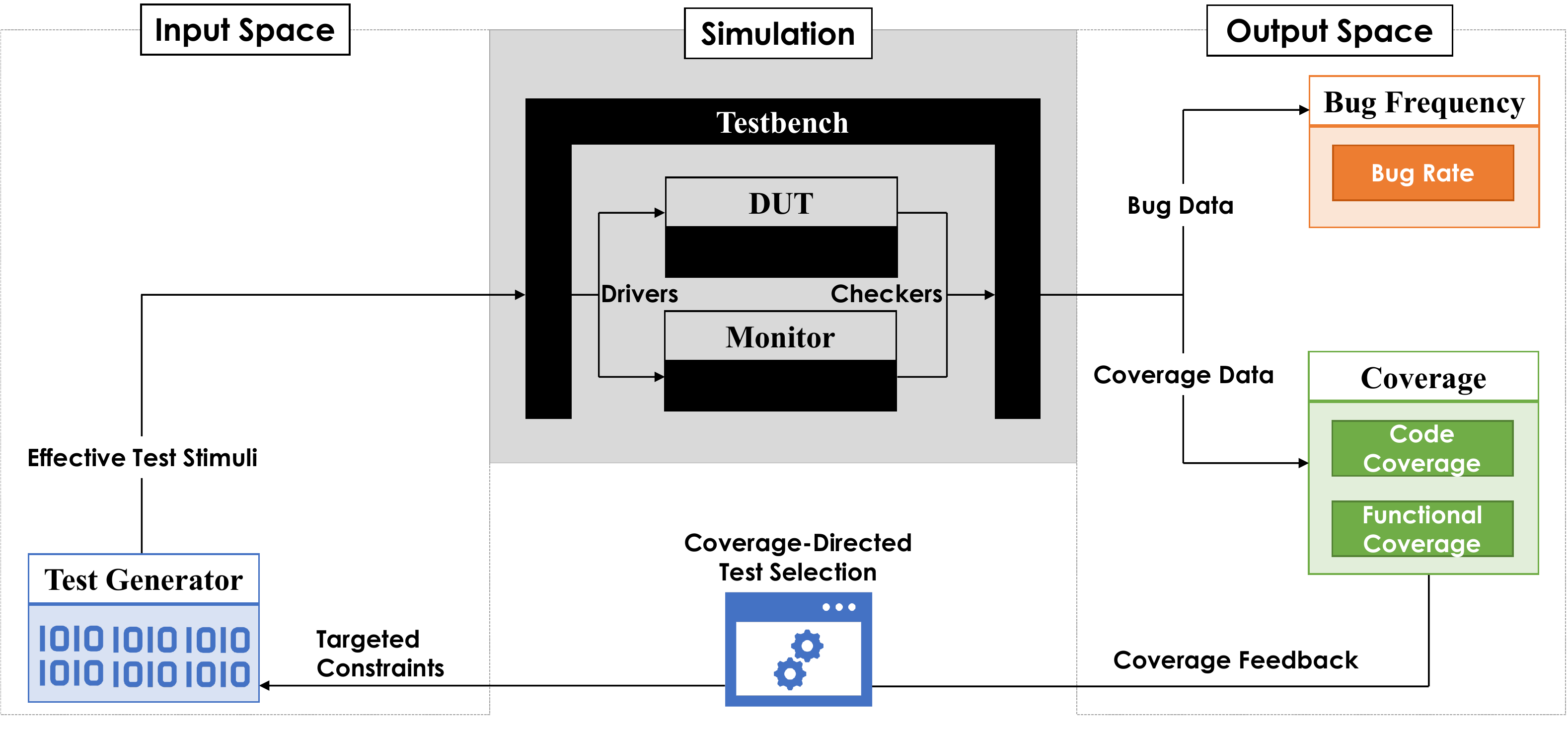}
    \caption{Coverage-Directed Test Selection}
    \label{fig:cds_only}
\end{figure}

When it comes to intelligent testing, verification engineers might aim to achieve two objectives aimed at exercising functionality of interest, as defined by the verification plan and the coverage model. The first objective is to identify and prioritise tests likely to trigger previously unexplored DUT functionality prior to simulation. Identifying and simulating diverse and dissimilar test stimuli is likely to result in  wider areas of DUT logic being triggered. The wider the spread of DUT logic triggered, the higher the chances of exposing hidden bugs and exercising new coverage. The redundancy inherent to constrained random testing is reduced because novel stimuli are less likely to repeatedly trigger the same DUT logic. Therefore, novel test prioritisation leads to redundancy reduction which, in turn, leads to efficiency gains during simulation as fewer tests are simulated and high coverage is achieved.

The second objective is to estimate, before simulation, the probability that a newly generated test will exercise one or more coverage holes by analysing data from previous simulations. Estimating this probability may enable us to predict which tests are most likely to exercise coverage holes before simulating them, and it provides us with a measure of how confident we can be about the prediction. Selecting and simulating tests that have a high probability of exercising coverage holes can boost coverage progress while increasing the likelihood of discovering bugs early. Testing is more effective because stimuli that exercise interesting and hard-to-reach functionality are given priority. A good indication of effective testing is a shifting of the coverage curve upwards and leftwards, meaning that higher coverage is being achieved quicker. When the horizontal axis represents the number of tests simulated, such a shift of the coverage curve is analogous to that exhibited when higher coverage is achieved with a smaller number of tests. The reduction in the number of simulated tests is therefore a suitable metric for both efficiency and efficacy experiments.

\subsection{Coverage-Directed Test Selection}\label{coverage_directed_test_generation}
Coverage-Directed Test Selection (CDS) is a method devised to automate the feedback loop from coverage analysis to test selection, depicted in Figure \ref{fig:cds_only}. It is adapted from feedback-based CDG \cite{IBM2003}, whose main goal is to produce constraints that automatically bias the test generator based on coverage feedback. The CDS engine uses supervised learning to model the relationship between coverage and test stimuli. Constraints are produced to automatically bias test selection towards stimuli that have the highest probability of exercising coverage holes. From this perspective, CDS automates the verification engineer's manual work of writing constraints by hand. CDS is flexible enough to be used with any binary classifier with minimal tuning, and can easily be extended to test generation problems. This reduces the effort required to encode domain knowledge into an AI model, and renders the method useful in contexts beyond test selection (although such contexts are beyond the scope of this paper).

More formally, let $T$ denote the set of previously simulated test stimuli derived from a constrained random test generator. For each test $t_i \in T$, there exists a set $X_i$ of $k$ constraints $X_i = \{x_1, x_2, \ldots, x_k\}$ that generated the test. Moreover, let $C$ represent the complete coverage model. After simulation of $T$, each coverage point $c \in C$ is assigned a binary value in $[0,1]$ depending on whether $c$ has not, or has, been exercised, respectively. Suppose that intelligent testing begins when we derive a set of simulation candidates $T'$ from the test generator.

The first task of CDS is to estimate the conditional probability $P$, of exercising a target coverage point $c_j \in C$, by generating a new test $t_j$, using a new set of constraints $X_j$. This conditional probability is written as shown in \eqref{eqn_conditional_prob} below. 
\begin{align}\label{eqn_conditional_prob}
    P(c_j | X_j, t_j)
\end{align}

To make the coverage space more manageable, the coverage model, $C$, is partitioned \cite{IBM2002} into a set, $G$, of $m$ non-overlapping coverage groups, as shown in \eqref{eqn_covgroups}. 
\begin{align}\label{eqn_covgroups}
    G = \{g_1, g_2, \ldots, g_m\} = C
\end{align}

Coverage groups are based on information from the verification plan. They enable model training and coverage analysis to be performed at the coverage group level. After simulation, the coverage groups $G$ are inspected to find target coverage groups that have been exercised by some minimum number of test stimuli, but still contain coverage holes. Each target coverage group, $g_j$, has its $n$ exercising stimuli, $T_{pos} \in T$, isolated and used as positive examples for training the group's classification model. For negative examples, $n$ stimuli, $T_{neg} \in T$, are randomly sampled from the tests that did not exercise any coverage within $g_j$. We train a supervised classification model for each target coverage group $g_j$ using tests $t_i$ in $T_{pos} + T_{neg}$ as training examples, constraints $X_i$ as training features for each $t_i$, and labels $[0,1]$ for each $t_i$ depending on whether or not it triggered coverage in $g_j$. 

The final CDS task is to classify a new test $t_j$ as one of two classes $Y = [1,0]$ depending on whether or not we expect coverage holes in target coverage group $g_j$ to be exercised by $t_j$, generated using constraints $X_j$. Applying a threshold, $\epsilon$, enables us to derive classes based on the computed conditional probability. A threshold of 0.5, for example, means that if $ P(c_j | X_j, t_j) > 0.5 $ then we predict $y = 1$. Otherwise, we predict $y = 0$. 

CDS has three major limitations. First, it is very difficult to learn a relationship between coverage and test stimuli. This is attributable to many reasons. Direct and well-understood relationships tend to be difficult to learn due the difference in abstraction levels between the coverage and test spaces. For example, in processor verification, test stimuli might be defined at the architectural level and generated in assembly language. Coverage tasks, on the other hand, are usually defined as cross-products \cite{IBM2002} based on micro-architectural requirements. Abstracting test stimuli further by representing them as tuples or vectors mitigates the issue, but the difficulty of learning meaningful relationships still persists. Alternatively, the relationship might not be learnable because the function to be approximated is too complex. The lack of positive training examples \cite{LCWang2019} is also problematic in CDS because, by aiming to exercise hard-to-reach and yet-uncovered DUT functionality, we are trying to make predictions about events that have occurred rarely, if at all, in the past. 

The second limitation of CDS is that it directly relies on learning from simulation traces. Simulation is expensive in terms of computation and simulation time, and selecting tests that exercise the hardest-to-reach coverage tasks might require at least tens of thousands of simulation cycles to collect adequate data to learn from. As observed in \cite{IBM2020}, learning from coverage feedback in such circumstances leads to resource inefficiencies linked to obtaining the required data. 

The third limitation of CDS arises from lack of a suitable distance metric. When targeting yet-unexercised coverage for which useful data are unavailable, an alternative approach is to relax the problem and learn from coverage points deemed `similar' to the unexercised coverage. This approach immediately raises challenges because a suitable distance metric (which clearly defines `similarity' in the coverage space) is unavailable in simulation-based verification. The traditional, syntax-based distance metrics such as Euclidean and Hamming are inherently limited in this context. While various approximations have been devised such as in \cite{IBM2009}, the lack of a semantically meaningful distance metric continues to be an issue in intelligent testing methodology. 

We will now show that NDV, which we examine below, is capable of addressing some of the limitations associated with CDS.

\begin{figure}[t]
    \centering
    \includegraphics[width=0.49\textwidth]{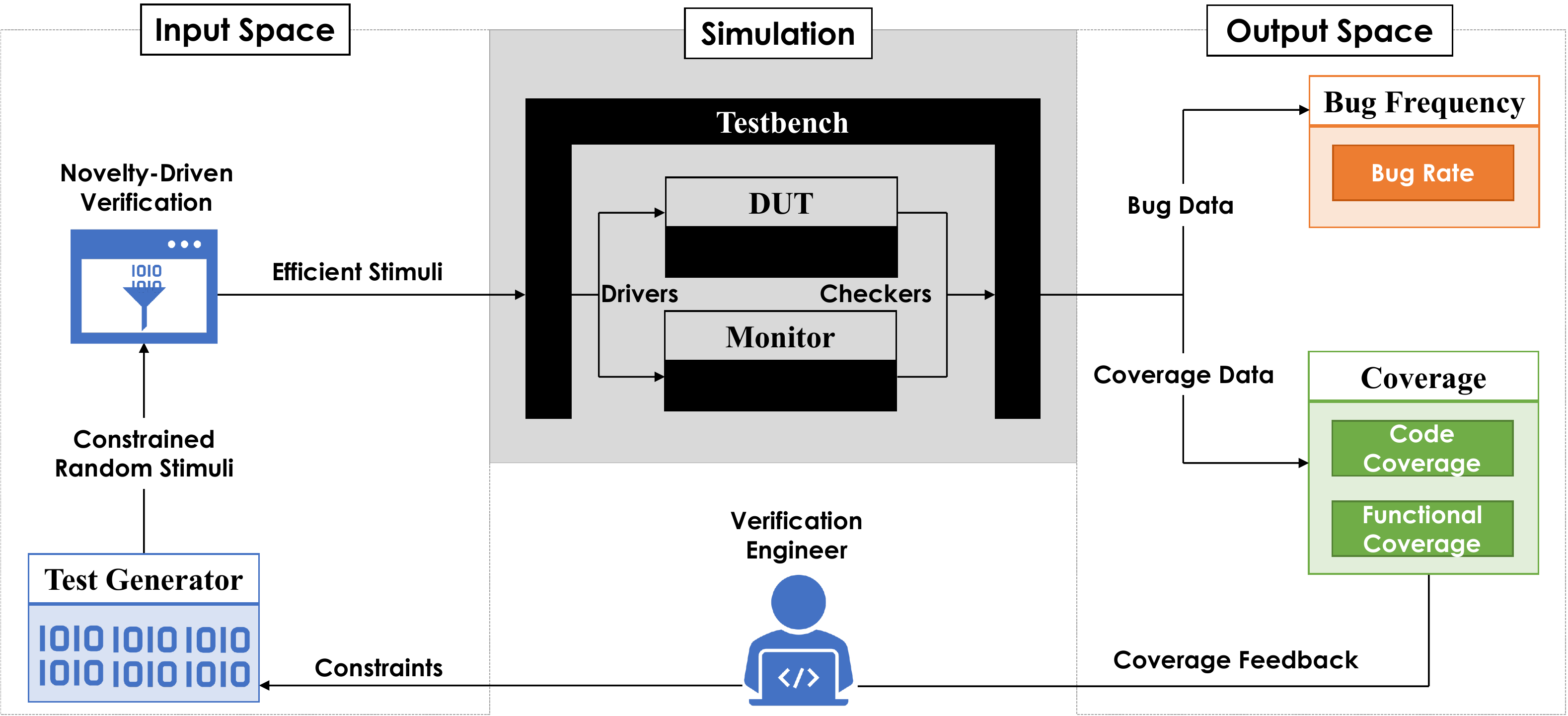}
    \caption{Novelty-Driven Verification}
    \label{fig:ndv_only}
\end{figure}

\subsection{Novelty-Driven Verification}
Novelty-Driven Verification (NDV) methods \cite{LCWang2012, Blackmoreetal2021} use novelty detection to identify and prioritise potentially useful tests for simulation. Our main assumption when applying NDV in simulation-based verification is as follows: tests that are novel in the input space are more likely to exercise novel DUT functionality, and correspondingly lead to novel coverage in the output space. NDV leads to a smaller set of test stimuli being simulated without the coverage suffering a corresponding reduction. The NDV process is depicted in Figure \ref{fig:ndv_only}. Similar to CDS, NDV requires that a stimuli generation source such as a constrained random test generator is available to create a large set of test stimuli as simulation candidates. The learning component operates on tests that have previously been simulated, by learning a model of similarity for these tests. When presented with newly generated test stimuli, the model ranks them based on their dissimilarity to previous stimuli. By identifying tests that are most dissimilar to those already simulated, then giving preference to those tests for simulation, there is a higher chance of exercising new DUT functionality and accelerating coverage progress.

The first task of NDV is to estimate a dissimilarity score, $\phi_{t_j}$, for each newly generated test $t_j \in T'$, relative to $T$. In this paper, this task is performed by training the One-Class Support Vector Machine (OCSVM) \cite{Scholkopf99OCSVM} unsupervised novelty detection algorithm using the set of previously simulated test stimuli $T$. 

The novelty detection algorithm builds a model of similarity $Z$, defined by a novelty threshold $z(T) = \theta$, where $\theta$ is a real number representing the decision boundary. The dissimilarity score $\phi_{t_j}$, for a given new test $t_j$, is computed as the signed distance of $t_j$ from the decision boundary, such that 
\begin{align}\label{eqn_dis_score_calc}
    \phi_{t_j} = \left(\sum_{t_i \in T}{\alpha_{t_i} L(t_j, T) - \theta}\right) \in (-\infty, +\infty)
\end{align}
In \eqref{eqn_dis_score_calc}, $\alpha_{t_i}$ is a parameter that determines the importance of each $t_i \in T$ in the model, and $L$ is a kernel function that defines the distance between $t_j$ and $T$. If $|\alpha_{t_i}| > 0$, then the test $t_i$ is considered a support vector, meaning that it is considered in the model calculation. If $|\alpha_{t_i}| = 0$, $t_i$ is considered a non-support vector and ignored. In general, the further away $t_j$ is from the decision boundary, the more negative $\phi_{t_j}$ is. The more negative $\phi_{t_j}$ is, the more novel $t_j$ is deemed to be relative to $T$. 

The second task of NDV is to take each test $t_j$'s dissimilarity score $\phi_{t_j}$ and use it to rank test stimuli based on decreasing dissimilarity, such that test $t_{jA}$ is selected for simulation before $t_{jB}$ if $\phi_{t_{jA}} < \phi_{t_{jB}}$. NDV can be extended to classify $t_j$ as one of two classes $Y = [-1,1]$ depending on whether $t_j$ is estimated to be novel or normal, respectively. The novelty threshold $\theta$ enables us to classify $t_j$ as being novel if $\phi_{t_j} < \theta$. Otherwise, $t_j$ is classed as being normal. 

NDV is computationally advantageous because it operates directly on test stimuli in the input space, without recourse to coverage data. NDV is unsupervised, simpler, and more `lightweight' than CDS.  Reducing the number of simulation candidates prior to simulation in a lightweight, unsupervised manner is extremely powerful when generation is cheap and simulation is expensive. Through NDV, we gain the ability to generate as many tests as our resources permit, but prioritise the most promising subset of those tests for simulation.

The main limitation of standalone NDV methods is that, by not being able to relate exercised coverage to their triggering test stimuli, and by solely relying on the stimuli data set to train an unsupervised model, the method likely loses a potentially valuable source of information. CDS can work in conjunction with NDV to mitigate this weakness by providing a coverage feedback component. However, the fusion of methods such as CDS and NDV in the literature is unexplored. Hence, we will now present a hybrid approach to intelligent testing incorporating both CDS and NDV.

\section{Hybrid Intelligent Testing Framework}
\label{hybrid_approach}

\begin{figure*}
    
    \centering
    \begin{subfigure}[t]{0.49\textwidth}
    \includegraphics[width=0.9\linewidth]{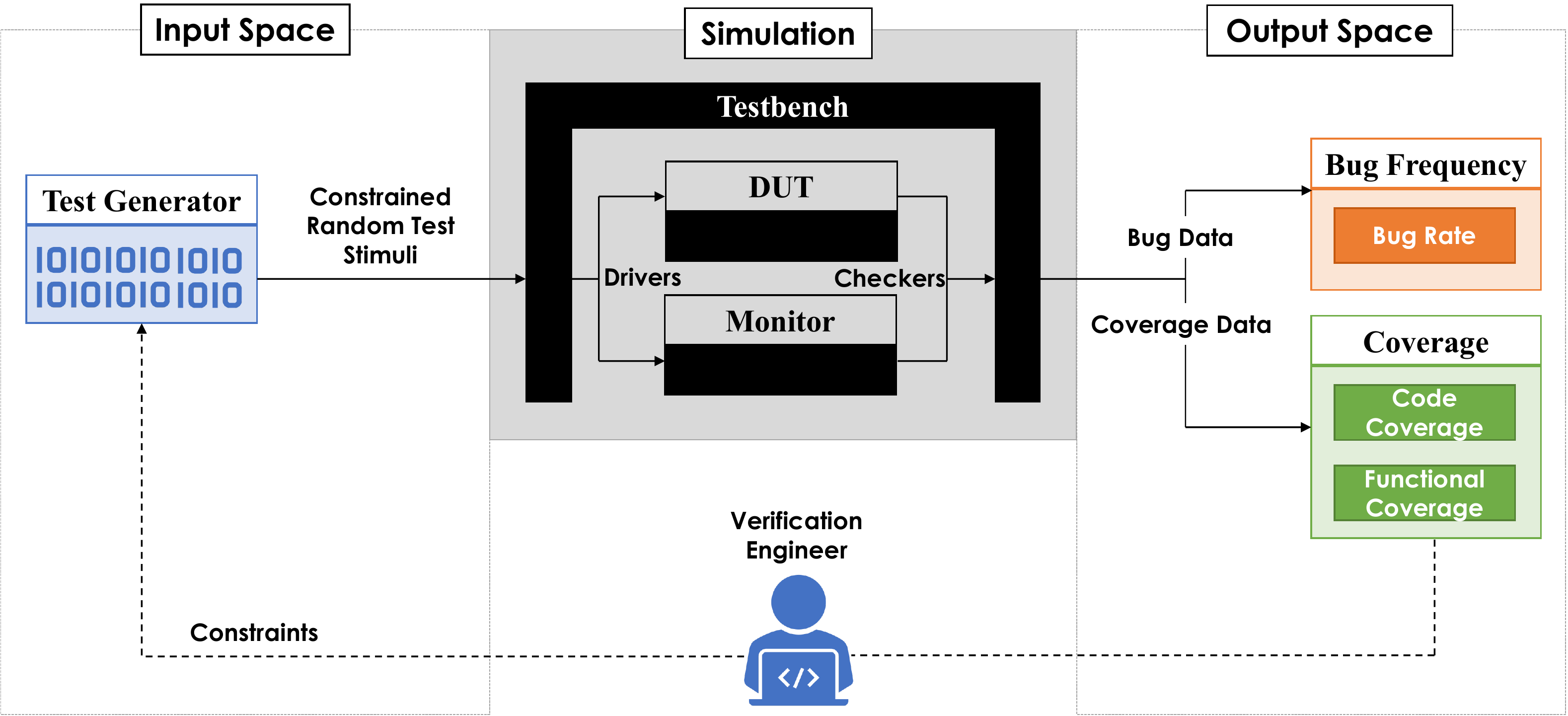} 
    \caption{Traditional Testing}
    \label{fig:traditional_testing}
    \end{subfigure}
    \hfill
    \begin{subfigure}[t]{0.49\textwidth}
    \includegraphics[width=0.9\linewidth]{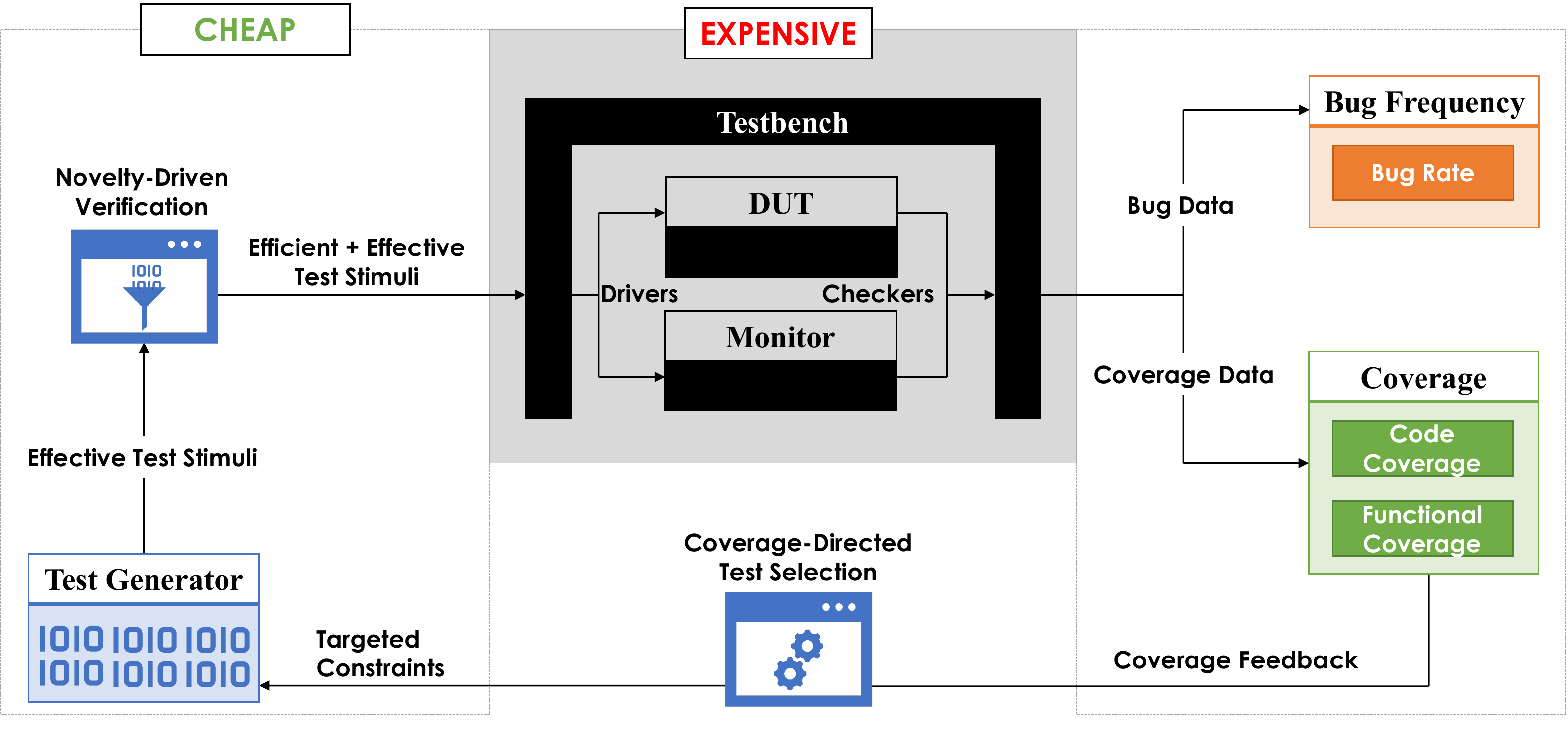}
    \caption{Intelligent Testing}
    \label{fig:intelligent_testing}
    \end{subfigure}
    
\caption{Traditional testing vs intelligent testing in simulation-based verification}
\label{fig:traditional_vs_intelli}

\end{figure*}

CDS methods automatically produce targeted constraints and effective tests specifically targeted at coverage closure, but suffer from issues largely stemming from supervised function approximation on simulation traces. On the other hand, NDV methods reduce redundancy without relying on supervised function approximation, but potentially lose useful information by not analysing coverage feedback. This suggests that CDS limitations can be addressed by NDV strengths, and vice-versa. Employing both CDS and NDV in a hybrid intelligent testing framework harnesses the strengths, while simultaneously limiting the drawbacks, of both CDS and NDV methods. An example of hybrid intelligent testing encompassing both CDS and NDV is depicted in Figure \ref{fig:traditional_vs_intelli}, where it is contrasted to `traditional' constrained random testing.

\subsection{Unified Hybrid Approach}
The unified hybrid approach to intelligent testing combines CDS and NDV in sequential order. When intelligent testing begins, only one of the methods is operating on stimuli at any given time. The unification procedure is as follows:
\begin{enumerate}
    \item Initialise simulation-based verification with traditional constrained random testing. Beginning with constrained random testing potentially exercises the DUT with a wide variety of randomly ordered tests, thereby increasing the likelihood of discovering hidden bugs that might not be easily found when targeting coverage.
    \item Switch from constrained random to intelligent testing to exercise coverage points of intermediate difficulty. Intelligent testing can be activated at a pre-determined point (e.g., when some level of functional coverage has been achieved), or when a given number of random stimuli have been simulated), or it can be activated dynamically (such as when functional coverage has not increased for a given number of iterations). The decision regarding which of the intelligent testing methods to activate first depends on the context in which the methods are being used. A good rule-of-thumb is to begin with NDV because it is unsupervised and requires less data to be operational. 
    \item Increase chances of exercising yet-uncovered coverage points by switching from one method of intelligent testing to the other. Step 3 is performed using CDS if NDV was used for step 2, and vice-versa. Similar to step 2 above, this switch can also be pre-determined or dynamic. Switching the intelligent testing method is crucial for step 3 because it `refreshes' the testing dynamic to one that has not been previously encountered, raising the chances of new coverage being exercised. For example, if NDV was used in step 2, coverage feedback was not taken into account. Switching from NDV to CDS means that coverage feedback starts being considered when selecting test stimuli, which inevitably changes the tests that will be simulated and increases the chances that new coverage will be exercised. The same reasoning works when CDS is used for step 2, and NDV for step 3. While the decision regarding which method to use in this step depends on context, a good rule-of-thumb is to use NDV in step 2 to gather as much data as possible for CDS supervised learning in step 3.
\end{enumerate}

\subsection{Intersected Hybrid Approach}
The intersected hybrid approach to intelligent testing combines CDS and NDV in parallel. When intelligent testing begins, both methods operate concurrently. The intersection procedure is as follows:
\begin{enumerate}
    \item Utilise constrained random testing during initial testing iterations.
    \item Switch from constrained random to intelligent testing. In each iteration, simulate test stimuli that meet both CDS and NDV criterion -- i.e., the intersection of CDS and NDV. The resulting test stimuli depend on the order of the intelligent testing operations, of which there are two possible configurations: (i) CDS-First, or (ii) NDV-First. The CDS-First pipeline first uses CDS to select stimuli that have the highest probability of exercising target coverage. The selected stimuli become inputs to a NDV process that selects the highest novelty scoring tests for simulation. An example of a CDS-First pipeline iteration is shown in Figure \ref{fig:intelligent_testing}. In a NDV-First pipeline, the order is reversed, with NDV being performed first, and the output of NDV being the input of CDS.
\end{enumerate}

\section{Experimental Setup}
\label{experiment}

We evaluated our unified approach to intelligent testing on an Infineon Technologies' Radar Signal Processing Unit (RSPU). The RSPU is a large, complex and highly configurable IP supporting the Advanced Driver Assistance Systems (ADAS) in Infineon's AURIX family of microcontrollers. The RSPU was previously used as the DUT in similar work that investigated an autoencoder-based approach \cite{Blackmoreetal2021}, and a one-class support vector machine-based approach \cite{Zheng2019} to NDV. However, it is worth noting that the coverage model and the training data used in this paper were different from those used in previous NDV work. We used a different coverage model largely containing hard-to-exercise whitebox coverage points. Our training data were subjected to less domain expertise-based feature engineering -- aside from encoding of categoricals and binning of large fields to reduce their cardinality -- to test the efficacy of this approach on data with minimal feature engineering. Therefore, the results can reasonably be expected to be different from those derived in previous NDV work.

Each test is represented as a tuple of approximately 300 fields. The metric we use to assess the success of these experiments is functional coverage. We use a functional coverage model containing about 6,000 whitebox cross-product coverage points. We frame CDS as a process of producing constraints that maximise our chances of selecting effective tests from the test database. With CDS, we learn which constraints have the highest probability of biasing new tests to exercise target coverage, then select the next iteration's tests based on their ability to satisfy learnt constraints. We frame NDV as a process of prioritising novel tests for simulation by ranking test stimuli in the test database and simulating them in order of decreasing dissimilarity. With NDV, we learn which tests have the highest novelty score (relative to previously simulated tests), then select the next iteration's tests in order of decreasing novelty.

Achieving 100\% functional coverage in real world verification requires approximately 2 million constrained random test stimuli. Because simulating each test takes an average of 2 hours, roughly 1,000 machines and EDA licences are consumed nearly continuously over the course of the RSPU's 6-month verification project. Additionally, several months of effort are spent by verification engineers writing constraints to target coverage holes. To enable rapid experimentation, a database was created to store each test and its coverage. A ranked regression of approximately 3,000 stimuli were found to enable coverage closure. These 3,000 stimuli and their coverage were added to the database for experimentation, along with 83,000 other constrained random test stimuli. This experimentation database enabled us to mimick simulation by querying for the coverage exercised by each simulated test, and reduced the total time required to run complete experiments from weeks to hours.

The NDV approach we use in this paper is similar to that employed by \cite{Zheng2019}. After initially sampling and simulating a number of constrained random tests, those tests are used to batch train a one-class support vector machine (OCSVM) \cite{Scholkopf99OCSVM} novelty detector. A dissimilarity score is subsequently obtained for all the remaining unsimulated tests. The unsimulated tests are then ranked for simulation according to dissimilarity score, with the most dissimilar tests being prioritised for simulation first.

Our CDS approach is implemented using a decision tree \cite{Breimanetal} algorithm. We adopt a rule-based approach to learning the relationships between test generation and coverage. As previously discussed, an immediate problem with applying CDS in practice is the lack of positive samples. By definition, we are trying to make predictions for events that are rare or have never happened. In other words, we cannot directly learn anything about coverage holes because there are no data to learn from. The approach we take to solve this problem is inspired by \cite{Ederetal07}, who proposed clustering the coverage space first. Because the clustering led to intuitive coverage groupings, it became possible to target a coverage hole in a given coverage group by tracing tests which exercise coverage points deemed close to the hole; inductively mining interpretable rules that describe those tests; and using the rules as constraints for generating new tests with the aim of covering the targeted hole. The lack of a well-defined distance metric also becomes apparent in CDS practice. The coverage space clustering approach in \cite{Ederetal07} seeks to learn from tests which have previously exercised coverage points deemed close to the hole. However, it is difficult to determine `closeness' without a well-defined distance metric. Coverage hole analysis \cite{IBM2002} suggests a user-defined partitioning of the coverage model instead. Partitioning divides the coverage space into manageable partitions based on readily available information, such as from the verification plan, in which coverage events tend to be grouped together based on shared traits such as their triggering DUT behaviour. By partitioning the coverage space \emph{apriori}, we derive meaningful coverage groups that enable us to find and target coverage holes, isolate stimuli which previously exercised neighbouring coverage points (within the same partition), and learn common rules that can be used as constraints for targeting those coverage holes. Partitioning also happens to be beneficial because mapping coverage groups to triggering stimuli requires less effort than attempting to map individual coverage points directly to generation constraints. The rules induced by the decision tree during during this mapping exercise can also be interpreted by verification engineers if required, making them useful in situations where constraint transparency is required.

\section{Results}
\label{results}

Hybrid intelligent testing experiments compared the performance of constrained random testing, CDS, NDV, against the Unified Hybrid Approach and the Intersected Hybrid Approach. The metric used to compare performance is the number of tests required by each method to achieve given levels of functional coverage. Because simulation is expensive in this context, reducing the number of tests simulated without sacrificing coverage is a desirable goal. The lower the number of tests required to reach a given functional coverage level, the more effective and efficient the intelligent testing method is considered to be. 

    \subsection{Standalone CDS and NDV}
Table \ref{savings_tbl_cds_ndv} shows the number of tests required to reach 95\%, 98\%, and 99\% functional coverage on the RSPU when comparing random selection to CDS and NDV as standalone methods. We regard 99\% as the most important of these functional coverage levels because, at that level, we reason that a verification engineer can intervene and manually close coverage on the remaining 60 coverage points. Random selection is the performance baseline against which the other selection methods are compared. The results show that NDV leads to the lowest number of simulations when compared to random selection, managing to save almost 16\% of tests at 99\% functional coverage. CDS required 6\% less tests than random selection to reach 99\% functional coverage. Although better than random, CDS savings were lower than NDV across all three functional coverage levels. There are two possible reasons for CDS underperformance: a lack of data to learn from (particularly during earlier iterations), and underfitting by the decision tree classifier. Both of the issues leading to CDS underperformance are addressed by the hybrid approaches.

    \subsection{Unified Hybrid Approach}
Table \ref{savings_tbl_uha} shows the number of tests required to reach different levels of functional coverage on the RSPU during unified hybrid approach (UHA) experiments. The table groups the results according to the intelligent test selection method performed first. To be clear, random selection initiates the verification process, and the switch from random selection to intelligent test selection occurs at 90\%. Stating that a method was performed first during UHA experiments means it was performed when functional coverage was between 90\% and 98\%. The last method to be performed as part of UHA experiments is performed between 98\% and 99\% functional coverage. 

The highest savings across all functional coverage levels were achieved when NDV was performed first during UHA experiments. The NDV-First UHA achieved 99\% functional coverage with 17\% less tests than random selection, and with almost 2\% less tests than standalone NDV. Such a result is to be expected due to two reasons. The first reason is that the performance of the NDV-First UHA at 95\%, and at 98\%, will be similar to the performance of standalone NDV. The second reason to expect this result is that it mitigates the problem of lack of quality supervised training data associated with CDS. NDV promotes test diversity, and the quantity of training data is not as much of an issue for this unsupervised method. As a result, NDV is extremely effective, even during earlier iterations. When the switch from NDV to CDS occurs at 98\%, the classifier immediately benefits from the relatively large quantities of data gathered during NDV. CDS-First UHA results show that reversing this process will not yield significant simulation savings. Therefore, NDV should be performed first in a UHA pipeline, thereby enabling data gathering for the final stage when CDS is performed.

    \subsection{Intersected Hybrid Approach}
Table \ref{savings_tbl_iha} shows the number of tests required to reach different levels of functional coverage on the RSPU during intersected hybrid approach (IHA) experiments. The table shows a high reduction in simulated tests regardless of the intelligent testing method that is used in the pipeline first. The CDS-First intersected hybrid approach is better at reducing simulations overall, reaching 99\% functional coverage with 18\% less tests than random selection, and 0.5\% less tests than the NDV-First IHA. At 99\% functional coverage, NDV-First IHA results show an improvement of 2\% over standalone NDV. 

The results in Table \ref{savings_tbl_iha} are a significant improvement for the standalone CDS method. This improvement can be attributed to the integration of NDV into the test selection pipeline. The decision tree classifier used to perform CDS is constrained, and it tends to underfit the data. Due to underfitting, the classifier can allocate the same high probability score to several tests, and consequently return them all as possible simulation candidates. During CDS-First IHA experiments, these simulation candidates would be piped as inputs to NDV. NDV allocates granular novelty scores to the simulation candidates, making it highly unlikely that no two novelty scores will be exactly the same. Novelty scores make it easy to objectively select 1 simulation candidate per target coverage group, instead of resorting to breaking ties randomly. In this way, the issue of CDS underfitting is mitigated by performing CDS followed by NDV in the same IHA pipeline.

The CDS-First IHA achieves lower simulation savings at 95\% functional coverage (23\%) compared to NDV-first IHA (27\%). This result is also likely to be caused by the lack of data typical in earlier stages of verification, although such lack of data seems to be mitigated by employing NDV as part of the IHA pipeline. CDS-First IHA performs better than NDV-First IHA when more data has been gathered, as evidenced by the higher savings achieved by CDS-First IHA at 98\% and 99\% functional coverage. Training data availability is clearly an important aspect to consider when deciding where to run CDS in an intelligent testing pipeline.

\begin{table}[t!]
\centering
\begin{tabular}{|c ||c |c |c |} 
 \hline
 \textbf{Functional Coverage} & \textbf{Random} & \textbf{CDS} & \textbf{NDV} \\ [0.5ex] 
 \hline
 95\% & 12866 & 11770 & 9049 \\
 \hline
 98\% & 29300 & 27179 & 22132 \\
 \hline
 99\% & 44200 & 41458 & 37227 \\
 \hline
 \textbf{Savings (vs. Random)} & & & \\
 \hline
 95\% &  & -8.52\% & -29.67\% \\
 \hline
 98\% &  & -7.24\% & -24.46\% \\
 \hline
 99\% & & -6.2\% & -15.78\% \\ [1ex] 
 \hline
\end{tabular}
\caption{Number of tests required to reach given levels of functional coverage -- Random vs CDS and NDV}
\label{savings_tbl_cds_ndv}
\end{table}

\begin{table}[t!]
\centering
\begin{tabular}{|c ||c |c |} 
 \hline
 \textbf{Functional Coverage} & \textbf{CDS-First UHA} & \textbf{NDV-First UHA} \\ [0.5ex] 
 \hline
 95\% & 11948 & 9294 \\
 \hline
 98\% & 27256 & 22036 \\
 \hline
 99\% & 41871 & 36601 \\
 \hline
 \textbf{Savings (vs. Random)} & & \\
 \hline
 95\% & -7.14\% & -27.76\% \\
 \hline
 98\% & -6.98\% & -24.79\% \\
 \hline
 99\% & -5.27\% & -17.19\% \\ [1ex] 
 \hline
\end{tabular}
\caption{Number of tests required to reach given levels of functional coverage -- Unified Hybrid Approach}
\label{savings_tbl_uha}
\end{table}

\begin{table}[t!]
\centering
\begin{tabular}{|c ||c |c |} 
 \hline
 \textbf{Functional Coverage} & \textbf{CDS-First IHA} & \textbf{NDV-First IHA} \\ [0.5ex] 
 \hline
 95\% & 9914 & 9373 \\
 \hline
 98\% & 21848 & 22038 \\
 \hline
 99\% & 36223 & 36440 \\
 \hline
 \textbf{Savings (vs. Random)} & & \\
 \hline
 95\% & -22.94\% & -27.15\% \\
 \hline
 98\% & -25.43\% & -24.78\% \\
 \hline
 99\% & -18.05\% & -17.56\% \\ [1ex] 
 \hline
\end{tabular}
\caption{Number of tests required to reach given levels of functional coverage -- Intersected Hybrid Approach}
\label{savings_tbl_iha}
\end{table}
	
\section{Conclusion}
\label{conclusion}

In this work, we have proposed incorporating coverage-directed test selection and novelty-driven verification methods into a hybrid intelligent testing framework for simulation-based verification. We have shown that enhancing constrained random test generation, through joint use of coverage-directed test selection and novelty-driven verification, yields benefits in automation and productivity by reducing verification resource consumption and accelerating coverage progress. Coverage-directed test selection biases test selection towards tests that plug coverage holes. Novelty-driven verification identifies novel test stimuli likely to exercise new functionality. Hybrid intelligent testing adds value through gains in testing effectiveness, and increased resource efficiency attributable to reducing test redundancy. These benefits are achieved by harnessing data which are already being gathered from the current functional verification workflow. 

The best performing intelligent testing method, intersected hybrid approach with coverage-directed selection being performed first, led to 99\% functional coverage being achieved with 18\% less tests than random selection. At production-scale RSPU verification, this would translate to approximately reducing simulation by 360,000 tests. This translates to 1.5 months of verification resource savings. We still have multiple ways of improving these intelligent testing methods, including feature selection, feature engineering, and dimensionality reduction. The results achieved by hybrid intelligent testing, combined with the potential for improvement, make this an encouraging research avenue to keep pursuing.

Both hybrid approaches were unable to significantly improve standalone novelty-driven verification in these experiments. This speaks to novelty-driven verification's ability to reduce simulation and drive coverage progress, with little tuning and computation. Novelty-driven verification performance is even more impressive considering the fact that coverage feedback is not taken into account. Coverage-directed test selection benefits more from the integration of novelty-driven verification, while the reverse is not necessarily true. We can conclude that focusing on coverage holes is not as good an accelerant of coverage progress as focusing on test diversity. However, novelty-driven verification does not produce constraints (which might be required to be human-readable by verification engineers). It is also difficult to estimate how novelty-driven verification will perform \emph{apriori}, whereas metrics such as classification accuracy could be used to estimate the future performance of coverage-directed test selection. We are currently conducting further research to develop this line of reasoning: how can we correlate learning metrics (e.g., accuracy, precision) to verification metrics (e.g. coverage, number of tests simulated)?

Future research will focus on optimising the performance of our hybrid intelligent testing framework. We intend to review the data used in the learning pipeline, and to tune the hyperparameters of our models in order to increase learning performance. We will also extend the techniques we used in this paper to other classification models such as neural networks, and to other novelty detection models such as isolation forests.

\bibliographystyle{IEEEtran}
\bibliography{main}

\end{document}